\documentstyle[12pt]{article}
\begin{document}

\newcommand{\siml}{\stackrel{<}{\sim}}
\newcommand{\simg}{\stackrel{>}{\sim}}
\baselineskip=1.2\baselineskip

\noindent
\begin{center}
{\large\bf
A Wavelet Analysis of Transient Spike Trains
of Hodgkin-Huxley Neurons
} 
\end{center}

\begin{center}
Hideo Hasegawa
\footnote{e-mail:  hasegawa@u-gakugei.ac.jp}
\end{center}

\begin{center}
{\it Department of Physics, Tokyo Gakugei University  \\
Koganei, Tokyo 184-8501, Japan}
\end{center}
\begin{center}
(Received \today)
\end{center}
\thispagestyle{myheadings}

\begin{center} 
{\bf Abstract}   \par
\end{center} 

\small
Transient spike trains consisting of $M$ (= 1 - 5) pulses generated 
by single Hodgkin-Huxley (HH) neurons, have been analyzed by using
both the continuous and discrete wavelet transformations (WT).
We have studied 
effects of variations in the interspike intervals (ISI) of input spikes
and effects of random noises
on the energy distribution and the wavelet entropy, which are expressed
in terms of the WT expansion coefficients.
The results obtained by the WT are discussed in connection 
with those obtained by the Fourier transformation.
\normalsize

\vspace{0.5cm}

\noindent
{\small\tt Keywords:  Wavelet transformation, Wavelet entropy ,
Hodgkin-Huxley neurons}
\vspace{1.0cm}

\section{Introduction}

During the last half century,
extensive experimental and theoretical studies have been made
on functions of brain where
neurons communicate information by action potentials
or spikes.
Because of the complexity of spike signals,
little is known how information is carried 
by spikes at the moment \cite{Rieke96}-\cite{Pouget00}.
It has been widely believed that information is encoded
in the average firing rate of individual neurons ({\it rate code}).
Andrian \cite{Andrian26} first noted the relationship
between neural firing rate and stimulus intensity, which
forms the basis of the rate code.
Actually firing activities of motor and sensory neurons are
reported to vary in response to applied stimuli.
In recent years, however, it has been proposed that detailed
spike timings play an important role in information
transmission ({\it temporal code}):
information is assumed to be encoded in interspike
intervals (ISIs) or in relative timings between
firing times of spikes \cite{Softky93}-\cite{Stevens98}.
Indeed, experimental evidences  have accumulated 
in the last several years, indicating a  use of
the temporal coding in neural systems \cite{Carr86}-\cite{Thorpe96}.
Human visual systems, for example, have shown to classify
patterns within 250 ms despite the fact that at least 
ten synaptic stages are involved from retina to 
the temporal brain \cite{Thorpe96}.
The transmission
times between
two successive stages of synaptic 
transmission are suggested to be no more than 10 ms 
on the average.
This period is too short to allow rates to be determined
accurately.

Although much of debates on the nature of the neural code has focused 
on rate versus temporal codes, there are other important
issue to consider: 
information is encoded in the activity of single (or very few) neurons
or that of a large number of neurons. 
A major question on the {\it population code} (or {\it ensemble code}) 
concerns how the information is carried
by correlation between different action potentials
fired by different neurons.
The simplest assumption is that little or no amount of information is
carried by such correlation and that information is coded
in the relative firing rates of the neurons
({\it ensemble rate code}).
This code model is supported by many experiments \cite{Geor86}
and has been adopted in the majority of analysis of 
neural coding \cite{Pouget00}\cite{Shadlen98}.
On the contrary, it is assumed that
relative timings between spikes in ensemble neurons
may be used as an encoding mechanism for perceptional
processing ({\it ensemble temporal code}) \cite{Hopfield95}-\cite{Rullen01}.
A number of experimental data supporting this code have been reported
in recent years \cite{Gray89}-\cite{Hatso98}.
For example, data have demonstrated that temporally
coordinated spikes can systematically signal sensory
object feature, even in the absence of changes
in firing rate of the spikes \cite{deCharms96}.
It is currently controversial what kind of code is employed
in the real neural systems \cite{Rieke96}-\cite{Pouget00}.

This shows the importance of studying how spike signals may convey
information with the sub-millisecond resolution.
The response of spikes to stimulus 
has been best studied when it is periodic in time, for which
the Fourier transformation (FT) method is usually adopted
for its analysis.
The FT decomposes a signal into its constituent components.
The information theory provides a powerful tool for
analyzing the nature and quality of a neuronal code.
A natural approach to quantify the degree of order of a complex 
signals is to consider its spectral entropy, as defined
from the FT power spectrum \cite{Powell79}.
The spectrum entropy shows how the FT power spectrum is
concentrated or widespread; an ordered activity with a narrow
peak in the frequency domain yields a low entropy
while a disordered activity with a widespread frequency
distribution leads to higher entropy.

The FT requires that a signal to be examined is stationary, 
not giving the time evolution of the frequency pattern. 
Actual biological signals are,
however, not necessarily stationary.
It has been reported that neurons in different regions have different
firing activities. 
Furthermore even within a given region, firing property
depends on its states.
The typical example is found in thalamus, which is the major gateway
for the flow of information toward the cerebral cortex.
In arousal spikes with gamma oscillations (30-70 Hz), mainly 40 Hz,
are reported, whereas spindle oscillations (7-14 Hz) and
slow oscillations (1-7 Hz) are found in early sleeping 
and deepen sleeping states, respectively \cite{Bal93}.
In hippocampus, 
gamma oscillation occurs in vivo, 
following sharp waves \cite{Freud96}.
In neo-cortex, gamma oscillation is observed under conditions of
sensory signal as well as during sleep \cite{Kisvarday93}.
Spike signals in cortical neurons are generally not stationary, 
rather they are transient signals or bursts \cite{Traub99}, 
whereas periodic spikes are found in systems 
such as auditory sytems of owl \cite{Sullivan98}
and the electrosensory system of electric fish \cite{Rose85}.

The limitation of the FT analysis can be partly resolved by using 
a short-time Fourier transform (STFT).  Assuming the signal
is quasi-stationary in the narrow time period,
the FT is applied with time-evolving narrow windows. 
Then STFT yields the time evolution of the frequency spectrum.
The STFT, however, has a critical limitation violating the
{\it uncertainty principle}, which asserts that if the window is too
narrow, the frequency resolution will be poor whereas
if the window is too wide, the time resolution will be less precise.
This limitation becomes serious for signals with much transient 
components, like spike signals.

The disadvantage of the STFT is overcome 
in the wavelet transform (WT) \cite{Astaf96}.
In contrast to the FT, the WT offers the two-dimensional expansion for
a time-dependent signal with the scale and translation parameters
which are interpreted physically as the inverse of frequency and
time, respectively. 
As a basis of the WT, we employ the {\it mother wavelet}
which is localized in both frequency and time domain.
The WT expansion is carried out in terms of a family of wavelets 
which is made by dilation and translation of the mother wavelet.
The time evolution of frequency pattern can be followed with an optimal
time-frequency resolution.

The WT appears to be an ideal tool for analyzing signals of 
a non-stationary nature.
In recent years the WT has been applied to an analysis of 
biological signals \cite{Samar99},
such as electoencephalographic (EEG) 
waves \cite{Blanco98}-\cite{Blanco96},
and spikes \cite{Hulata00}-\cite{Strat01}.
EEG is a reflection of the activity of ensembles of neurons 
producing oscillations. By using the WT, we
can obtain the time-dependent decomposition of
EEG signals to $\delta$ (0.3-3.5 Hz), 
$\theta$ (3.5-7.5 Hz), $\alpha$ (7.5-12.5 Hz),
$\beta$ (12.5-30.0 Hz) and $\gamma$ (30-70 Hz) 
components \cite{Blanco98}-\cite{Blanco96}.
It has been shown that the WT is a powerful tool to the spike sorting
in which coherent signals of a single target neuron 
are extracted from mixture
of response signals \cite{Hulata00}-\cite{Zour97}. 
The WT has been probed \cite{Hulata00}  
to be superior than the conventional
analysis methods like the principal component 
analysis (PCA) \cite{Oja95}.

Besides the FT, another useful way to describe the dynamical behavior
of stationary signals is to
use Lyapunov exponents and correlation dimensions \cite{Grassberger83}.
Lyapunov exponents, which were initially introduced
for an analysis of chaos, measure the rate at which nearby points
on an attractor diverge or converge along nearby trajectories.
Correlation dimensions are meaningful quantities to examine
whether a given signal is chaotic or not.
We must note, however, that these quantities are defined only for
stationary signals.

It is the purpose of the present paper to make a WT analysis
of {\it transient} spike signals which have not been
quantitatively discussed so far.
We have analyzed spike trains consisting of $M$ (=1-5) pulses,
generated by the Hodgkin-Huxley (HH) neuron model \cite{Hodgkin52}.
We have calculated
the energy distribution and the wavelet entropy which are expressed 
in terms of the WT expansion coefficients.
The WT is classified to the continuous wavelet 
transform (CWT) and the discrete wavelet transform (DWT).
The former treats the scale and translation parameters 
as continuous variables
while the latter adopts the variables only at the discrete points. 
Both the CWT and DWT have advantages.
The CWT provides us with the intuitively clear results.
On the contrary, the DWT, which is based on the ortho-normal
basis, can be quickly performed by using the multi-resolution analysis.

Our paper is organized as follows:
In the next $\S$ 2.1, we describe the 
HH neuron model \cite{Hodgkin52}, 
which is considered to be the most realistic neuron model
among proposed theoretical models.
In $\S$ 2.2. we briefly mention the CWT and DWT,
presenting expressions for the energy distribution
and the wavelet entropy. 
The results of WT analysis of spike signals by using both CWT 
and DWT 
are presented in $\S$ 3, where
the effect of ISI of spike signals and
of noises are studied.
The final $\S$ 4 is devoted to discussions and conclusion.

\section{Calculation Method}

\subsection{HH Neuron Model}

In order to generate spike trains,
we adopt a single HH model \cite{Hodgkin52}.
The neuron is assumed to be in the environment with 
the independent Ornstein-Uhlenbeck (OU) noises.
Dynamics of the membrane potential $V$ of
the HH neuron
is described by the non-linear differential
equation given by 

\begin{equation}
\bar{C} \:d V(t)/d t = -I_{\rm ion}(V, m, h, n)  
+ I_{\rm ps}+ I_{n},
\end{equation}
where $\bar{C} = 1 \; \mu {\rm F/cm}^2$ is the capacity of the membrane.
The first term of Eq.(1) expresses the ion current given by
\begin{equation}
I_{\rm ion}(V, m, h, n) 
= g_{\rm Na} m^3 h (V - V_{\rm Na})
+ g_{\rm K} n^4 (V - V_{\rm K}) 
+ g_{\rm L} (V - V_{\rm L}).
\end{equation}
Here the maximum values of conductivities 
of Na and K channels and leakage are
$g_{\rm Na} = 120 \; {\rm mS/cm}^2$, 
$g_{\rm K} = 36 \; {\rm mS/cm}^2$ and
$g_{\rm L} = 0.3 \; {\rm mS/cm}^2$, respectively; 
the respective reversal potentials are   
$V_{\rm Na} = 50$ mV, $V_{\rm K} = -77$ mV and 
$V_{\rm L} = -54.5 $ mV.
Dynamics of the gating variables of Na and
K channels, $m, h$ and $n$,
are described by the ordinary differential equations,
whose details have been given elsewhere \cite{Hasegawa00}.

The second term in Eq.(1) denotes the postsynaptic currents given by
\begin{equation}
I_{\rm ps} = \;
g_{s}\:(V_a - V_s) \:\alpha(t-t_{im}),
\end{equation}
with the alpha function $\alpha(t)$:
\begin{equation}
\alpha(t) = (t/\tau_{\rm s}) \; e^{-t/\tau_{\rm s}} \:  \Theta(t),
\end{equation}
where $\Theta (t)=1$ for $x \geq 0$ and 0 for $x < 0$.
Equation (3) expresses the postsynaptic current 
of the neuron, which is
induced by the  presynaptic spike-train input given by

\begin{equation}
U_i(t) = V_{\rm a} \: \sum_{m=1}^M  
\: \delta (t - t_{{\rm i}m}).
\end{equation}
where $V_a$ means its magnitude, $M$ the number of input pulses
and $t_{{\rm i}m}$ is the $m$-th firing time of the input.
The interspike interval (ISI) of input signal is defined by
\begin{equation}
T_{im} = t_{{\rm i}m+1} - t_{{\rm i}m}.
\end{equation}

The third term in Eq.(1) denotes the OU noise given by
\begin{equation}
\tau_{n}\: d I_{n}/d t = - I_{n} + \xi(t),
\end{equation}
with the Gaussian white noise $\xi(t)$:
\begin{equation}
< \overline{\xi(t)}> = 0, 
\end{equation}
\begin{equation}
<\overline{\xi(t) \:\xi(t')}> = 2 D \:\delta(t-t'),
\end{equation}
where the bracket $ < X >$ denote the time average.
The intensity and the correlation time of white noises are
given by $D$ and $\tau_n$, respectively

Differential equations given by Eqs.(1)-(9)
are solved by the forth-order Runge-Kutta method
by the integration time step of 0.01 ms
with double precision.
Some results are examined by using
the exponential method.
We incorporate OU noises given by Eqs.(7)-(9)
using the method
of Fox, Gatland, Roy and Vemuri \cite{Fox88}.
The initial conditions for the variables are given by
\begin{equation}
V(t)= -65 \:\: \mbox{\rm mV}, m(t)=0.0526,  
h(t)=0.600, 
n(t)=0.313, \:\: \mbox{\rm for} 
\:\:  \mbox{at} \: t=0,
\end{equation}
which are the rest-state solution of a single
HH neuron.
We adopted parameters of
$V_a=30$, $V_c=-50$ mV, and
$\tau_{\rm s} = \tau_{n}=2$ ms,
with which the HH neuron is excitable receiving suprathreshold inputs. 

\subsection{Wavelet Analysis}

\subsubsection{Continuous Wavelet Transformation}

The continuous wavelet transformation (CWT) for a given
regular function
$f(t)$ with a vanishing average is defined by
\begin{equation}
c(a, b) = \int {\rm d}t  \; \psi_{a,b}^{*}(t) \;f(t) 
\equiv <\psi_{a,b}(t),\: f(t)>,
\end{equation}
where $a$ and $b$ express the scale change and translation,
respectively, and the star denotes the complex conjugate.
The wavelet function $\psi_{a,b}(t)$ is generated by dilation
and translation of
the {\it mother wavelet} $\psi(t)$ such as
\begin{equation}
\psi_{a,b}(t) = \mid a \mid^{-1/2} \psi(\frac{t-b}{a}). 
\end{equation}
The parameters of $a$ and $b$ in Eqs.(11) and (12)
stand for the inverse of
the frequency and the time, respectively.
Then the CWT transforms the time-dependent function $f(t)$
into the frequency- and time-dependent function $c(a,b)$. 
The mother wavelet is a smooth function with good localization
in both frequency and time spaces.
A wavelet family given by Eq.(12) play a role of elementary
function, representing the function $f(t)$ as a superposition
of wavelets $\psi_{a,b}(t)$.

The {\it inverse} of the wavelet transformation may be given by
\begin{equation}
f(t) = C_{\psi}^{-1}  \int \: \frac{{\rm d}a}{a^2} \int {\rm d}b \;
c(a, b) \;\psi_{a,b}(t),
\end{equation}
when the mother wavelet satisfies the following two conditions:

\noindent
(i) the admissibility 
condition given by
\begin{equation}
0 < C_{\psi} < \infty,
\end{equation}
with
\begin{equation}
C_{\psi} = 
\int_{-\infty}^{\infty} {\rm d}\omega \mid \hat{\Psi}(\omega) \mid^2/
\mid \omega \mid, 
\end{equation}
where $\hat{\Psi}(\omega) $ is the Fourier
transform of $\psi(t)$,
and 

\noindent
(ii) the zero mean of the mother wavelet:
\begin{equation}
\int_{-\infty}^{\infty}  {\rm d}t \; \psi(t) = \hat{\Psi}(0) = 0.
\end{equation}

\vspace{0.5cm}

\noindent
{\it Energy Distribution}

Parseval's theorem shows that the power spectrum is given by
\begin{equation}
E_{\rm tot} = \int {\rm d}t \mid f(t) \mid^2 
= \: \int \frac{{\rm d}a}{a^2} \; \int {\rm d}b \; \epsilon(a,b) \;
= \: \int \frac{{\rm d}a}{a^2} \; E(a), 
\end{equation}
where the energy-density distribution $\epsilon(a,b)$ and 
the $a$-dependent energy is given by
\begin{equation}
\epsilon(a,b) = C_{\psi}^{-1} \mid c(a, b) \mid^2,
\end{equation}
\begin{equation}
E(a) =  \: \int {\rm d}b \; \epsilon(a,b). 
\end{equation}

\vspace{0.5cm}

\noindent
{\it Wavelet Entropy}

Equation (17) suggests that the probability of the power spectrum
relevant to the states of $a$ and $b$ is given by
\begin{equation}
p(a, b) = \epsilon(a, b)/E_{\rm tot},
\end{equation}
with the normalization condition:
\begin{equation}
\: \int  \frac{{\rm d}a}{a^2}\; \int {\rm d}b \;p(a, b) = 1.
\end{equation}
We may define the wavelet entropy given by
\begin{equation}
S = \: \int  \frac{{\rm d}a}{a^2}\; \int {\rm d}b \;s(a, b)
= \: \int \frac{{\rm d}a}{a^2} \; S_1(a),
\end{equation}
where $s(a, b)$ and $S_1(a)$ are given by
\begin{equation}
s(a, b)= -\: p(a, b) \; {\rm log}_2[p(a, b)],
\end{equation}
\begin{equation}
S_1(a)= \: \int \; {\rm d}b \; s(a ,b),
\end{equation}

Alternatively, we may define the wavelet entropy by
\begin{equation}
S' = \: \int {\rm d}a \; S_2(a),
\end{equation}
with
\begin{equation}
S_2(a)= - \; q(a) \; {\rm log}_2[q(a)],
\end{equation}
where the probability $q(a)$ is given by
\begin{equation}
q(a)=  a^{-2}\;E(a)/E_{\rm tot},
\end{equation}
satisfying the normalization condition:
\begin{equation}
\int \; {\rm d}a \; q(a) = 1,
\end{equation}
the prefactor $a^{-2}$ in Eq.(27)
arising from the space factor relevant to the integration
with respect to the variable $a$. 
Our numerical calculations to be presented in the following subsection
shows that $S_1(a)/a^2$ defined by Eq.(24) yields similar results
as $S_2(a)$ defined by Eq.(26)

\vspace{0.5cm}

\subsubsection{Discrete Wavelet Transformation}

The discrete wavelet transformation (DWT) is defined
for {\it discrete} values of $a=2^{j}$ and $b=2^{j}k$ 
($j,k$: integers) as
\begin{equation}
c_{jk} \equiv c(2^{j}, 2^{j}k) = <\psi_{jk}(t),\; f(t)>, 
\end{equation}
with
\begin{equation}
\psi_{jk}(t) = \mid 2 \mid^{-j/2} \psi(2^{-j} t - k).
\end{equation}
The ortho-normal condition for the wavelet functions is
given by
\begin{equation}
<\psi_{jk}(t), \: \psi_{j'k'}(t)> 
= \delta_{jj'}\:\delta_{kk'},
\end{equation}
which leads to the inverse DWT:
\begin{equation}
f(t) = \sum_{j} \; \sum_{k} \;c_{jk} \;\psi_{jk}(t).
\end{equation}

In the multiresolution analysis (MRA) 
of the DWT, we introduce a scaling function
$\phi(t)$, which satisfies the recurrent relation with $2K$
masking coefficients, $h_{k}$, given by
\begin{equation}
\phi(t) = \surd 2\sum_{k=0}^{2K-1} h_k \;\phi(2t- k),
\end{equation}
with the normalization condition for $\phi(t)$ given by
\begin{equation}
\int {\rm dt} \; \phi(t) = 1.
\end{equation}
A family of wavelet functions are generated by
\begin{equation}
\psi(t) = \surd 2\sum_{k=0}^{2K-1} (-1)^k h_{2K-1-k} \;\phi(2t- k).
\end{equation}
The scaling and wavelet functions satisfy the orthogonal
relations:
\begin{equation}
<\phi(t), \phi(t-m)> = \delta_{m0}, \\
<\psi(t), \psi(t-m)> = \delta_{m0}, \mbox{and}  \\
<\phi(t), \psi(t-m)> = 0.
\end{equation}
The set of masking coefficients $h_j$ 
are chosen so as to satisfy the conditions shown above.

The simplest wavelet function for $K=1$ is the Harr wavelet
for which we get $h_0 = h_1 = 1/\surd 2$ and 
\begin{eqnarray}
\psi_{\rm H}(t) &=& 1 \;\;\mbox{for $0 \leq t < 1/2$} \\
&=& -1 \;\;\mbox{for $1/2 \leq t < 1$}, \\
&=& 0 \;\;\mbox{otherwise}.
\end{eqnarray}
In the more sophisticated wavelets like
the Daubechies wavelet, an additional condition given by
\begin{equation}
\int {\rm d}t\; t^{\ell} \;\psi(t) = 0, \;\;\; 
\mbox{for $\ell=0,1,2,3.....L-1$}
\end{equation}
is imposed for the smoothness of the wavelet function.
Furthermore,
in the Coiflet wavelet, for example, a similar smoothing 
condition is imposed also for the scaling function as
\begin{equation}
\int {\rm d}t\; t^{\ell} \;\phi(t) = 0, \;\;\; 
\mbox{for $\ell=1,2,3.....L'-1$}
\end{equation}

In principle the expansion coefficients $c_{jk}$ in DWT 
may be calculated 
by using Eq.(29) for a given function $f(t)$ and
an adopted mother wavelet $\psi(t)$. 
This integration is, however, inconvenient,
and in an actual fast wavelet transformation, 
the expansion coefficients are obtained 
by a matrix multiplication with the use of the iterative formulae
given by the masking coefficients and 
expansion coefficients of the neighboring levels of indices,
$j$ and $k$ \cite{Astaf96}.

\vspace{0.5cm}
\noindent
{\it Energy Distribution}

Parseval's theorem for the DWT is given by
\begin{equation}
E_{\rm tot} = \int {\rm d}t \mid f(t) \mid^2
= \sum_j \sum_k \epsilon_{jk}
= \sum_j E_j, 
\end{equation}
where $\epsilon_{jk}$, the energy distribution specified 
by $j$ and $k$, and
the $j$-dependent energy $E_j$ are given by
\begin{equation}
\epsilon_{jk} = \mid c_{jk} \mid^2,
\end{equation}
\begin{equation}
E_j = \sum_k \epsilon_{jk}.
\end{equation}

\vspace{0.5cm}

\noindent
{\it Wavelet Entropy}

We may define the wavelet entropy as
\begin{equation}
S = \sum_j S_{j},
\end{equation}
with
\begin{equation}
S_j = - \sum_k \;p_{jk} \; {\rm log}_2 \;p_{jk},
\end{equation}
where the probability $p_{jk}$ is given by
\begin{equation}
p_{jk} = \epsilon_{jk}/E_{\rm tot},
\end{equation}
satisfying the normalization condition:
\begin{equation}
\sum_j \; \sum_k \;p_{jk} = 1.
\end{equation}

An alternative definition of the entropy is 
given by \cite{Rosso01}\cite{Blanco98}
\begin{equation}
S' = \sum_j S_{j}',
\end{equation}
with
\begin{equation}
S_j^{'}= - \;q_j \; {\rm log}_2 \; q_j,
\end{equation}
where $q_j$ expresses the probability given by
\begin{equation}
q_j = E_j/E_{\rm tot},
\end{equation}
with the normalization condition:
\begin{equation}
\sum_j \; q_j = 1.
\end{equation}
A comparison between $S_j$ and $S_j^{'}$ will be numerically made
in the following section.

Hereafter, time, voltage,
current, noise intensity ($D$), energy ($E_{\rm tot}$) and entropy ($S$)
are expressed in units of ms, mV, $\mu {\rm A/cm}^2$,
$\mu {\rm A}^2/{\rm cm}^4$, 
${\rm mV}^2\cdot {\rm ms}$ and bits, respectively.


\section{Calculated Results}


Clustered spike trains to be analyzed are generated 
by using the HH neuron
which receives input signals consisting of 
$M$ (=1-5) pulses with the ISI of $T_{im} =25$ ms ($m$=1-4).
Pulse clusters are separated by the longer interval of 100 ms.
We choose this ISI value of 25 ms
because $\gamma$-wave spikes with the average frequency
of about 40 Hz are ubiquitous 
in brain \cite{Bal93}-\cite{Kisvarday93}.
The time dependence of $U_{\rm i}$,
input potential $I_{\rm} \;(=I_{\rm ps} + I_{\rm n})$ 
and membrane potential $V$ is shown in Fig. 1.
Spikes fire after an injection
of input pulses with a delay of about 2 $\sim$ 3 ms.

Generated spike trains have been analyzed with the use of the WT.
One of the advantages of the WT over the FT is that we can choose
a proper mother wavelet depending on the shape of signals
to be examined.
Among many candidates of mother wavelets, we have adopted 
the Coiflet because its shape is similar to that of spikes.
Compromising the accuracy and the computation effort,
we have decided to adopt the Coiflet of order 3 as the mother wavelet
for our analysis.

We analyze the time-dependent membrane potential $V(t)$
by setting the signal to be $f(t) = V(t) - <V(t)>$
in Eqs.(11) and (29), $<V(t)>$ being the time average of $V(t)$. 
Spike signals $V(t)$ are assumed to be given at the sampled $t$ values
with a sampling time of $\Delta t = 1$ ms, and are
analyzed by both the CWT and DWT with the use of
MATLAB wavelet tool box.

The analyzed spike train is depicted in Fig. 2(a) and
the pattern of the expansion
coefficients $c(a, b)$ obtained by the CWT is shown 
in the $(a, b)$ space of Fig. 2(b), where
the medium dark tone denotes the zero level and the black (white)
expresses the negative (positive) value. 
Note that $1/a$ and $b$ physically 
stand for the frequency and time, respectively.
The pattern of the energy-density distribution  
$\epsilon(a,b)$ [$= \mid c(a, b)\mid^2$ by setting $C_{\psi}=1$ 
in Eq.(18) hereafter]  
is shown in Fig. 2(c), where black and white
denote the minimum (zero) and maximum levels, respectively.
Patterns near the both sides along vertical axis in Fig. 2(b) and (c) 
are due to the {\it edge effect} and should not be taken seriously.
Figure 2(c) shows that
for a single ($M=1$) spike signal, the energy-density pattern shows a cone
which is progressively broader at larger $a$.
For spike signals with multiple $M$, cones of spikes are
separable for small $a$ whereas they merge for larger $a$, yielding
intrigue structure in the pattern.
Figure 3 shows the $b$ dependence of the expansion coefficients 
$c(a, b)$ for $a$ = 2, 4, 8, 16, 32 and 64. 
We notice that the CWT coefficients for smaller $a$ oscillates
with shorter period.
The magnitudes of $c(a,b)$ are largest for $a \sim 8-16$ n Fig. 3: note
vertical scales for $a$= 2 and 64 to be different from those for the
other $a$ values. 

In what follows, we pay our attention to the $M=3$ spike in order
to perform more detailed analysis.
Figure 4(a) shows the $M=3$ spike signal to be analyzed. 
The energy-density distribution 
analyzed by the CWT is depicted in Fig. 4(b) .
At small $a$ each pulse yields a characteristic cone pattern.
At $10 \siml a \siml 20$, cones arising the consecutive pulses
separated by $T_o$ = 25 msec, begin to merge.
At larger $a \simg 40$ cones originating from
the first and the last pulses, which are separated by 
$2 T_o$ = 50 msec, begins to correlate.
The pattern has structures near both end sides
of vertical axis because of the edge effect.

Figure 5 expresses the $b$ dependence of the expansion 
coefficients of $c(a, b)$ for typical $a$ values
calculated in the CWT.
As noticed in Fig. 3, the magnitudes of $c(a, b)$ is most
significant for $a \sim 16$.
This is more clearly seen in the $a$-dependence of the energy-density
distribution $E(a)$
shown in Fig. 6(a) which has a peak at $a \sim 20$.
However, when taking into account 
the space factor $a^{-2}$ in the integration
with respect to the variable $a$ [see Eq.(17)], 
we note that $E(a)/a^2$
has a peak at much a lower value of $a=4$.

The $a$-dependent average [$\mu(a)$] and
root-mean-square value [$\sigma(a)$] of CWT coefficients
are plotted in Fig. 7(a) and (b), respectively.
A rise in $\mu(a)$ at $a > 30$ is due to the edge effect.
We note that $\sigma(a)$ has a peak at $a \sim 20$.
When the space factor $a^{-2}$ is included, the peak position of 
$\sigma(a)/a^2$ moves to the lower value of $a=4$.

So far we have employed the CWT. When we apply the DWT
to the signal shown in Fig. 4(a),
we get the $j-$ and $k$-dependent
energy distribution, $\epsilon_{jk}$, which 
is plotted in Fig. 4(c).
Because the scale ($a=2^j$) and translation ($b=2^j \:k$) parameters
are discrete in the DWT, the energy distribution, $\epsilon_{jk}$,
defined by Eq.(43) are depicted by the blocks in the $(a, b)$ space.
The DWT decomposition of the spike signal:
$V= \sum_{j=1}^{6} a_j + d_6$, is shown in Fig. 8. A contribution
from $j=2$ ($a=4$) to $V$ seems to be predominant.
This is supported by the calculated $j$-dependent energy 
$E_j$ plotted 
in Fig. 6(c), which has the maximum at $j=2$ ($a=4$).
The $j$ dependence of the average ($\mu_j$) 
and RMS value ($\sigma_j$) of the DWT coefficients
are plotted in Fig. 7(c) and (d), respectively.
The maximum in $\sigma_j$ is realized at $2^j=8$ ($j=3$),
which should be compared with the
maximum at $a=4$ in $\sigma(a)/a^2$.
These comparisons show that the $a$ dependence of
$E(a)/a^2$ [$\sigma(a)/a^2$] in the CWT are in fairly
good agreement with the $j$ dependence of $E_j$ [$\sigma_j$]
in the DWT.

Next we discuss the wavelet entropy calculated
in terms of the WT expansion coefficients.
Figure  9(a) shows the contour map depicting the
entropy density $s(a, b)$ defined by Eq.(23) with the CWT.
Figure  9(b) shows the $a$ dependence of $s(a,b)$
for typical values of $a$ = 103, 109 and 115,
which are indicated by vertical, dashed lines in Fig. 9(a).
$s(a, b)$ has peaks at $a \sim 10$, $a \sim 15$ and $a \sim 20$
for $b$ = 103, $b$ = 109 and $b$ = 115, respectively.
Figure  9(c) expresses the $b$ dependence of $s(a,b)$
for $b$ = 10, 20 and 30 ms, which are shown by horizontal, dashed
lines in Fig. 9(a).
Although $s(a, b)$ for $a=10$ has the maximum peaks
at the firing times $t_{{\rm o}n}$ (=103, 128 and 153 ms), 
it has also side-peaks before and
after  $t_{{\rm o}n}$.
For $a=20$ and 30, s(a, b) has peaks not only at $t_{{\rm o}n}$
but also between the firing times.
The $a$ dependence of the CWT entropy $S_1(a)$ defined by Eq.(24) 
is plotted in Fig. 6(b), which has a peak at
$a=20$, just as $E(a)$ shown in Fig. 6(a).
We note, however, that $S_1(a)/a^2$ has a peak at a lower 
value of $a =3$, which may be compared with 
the peak position at $a=4$
of the entropy
$S_2(a)$ defined by Eq.(26).

The DWT entropies, $S_j$ and $S'_j$, defined 
by Eqs.(46) and (50),
are shown in Fig. 6(d).
Although $S_j$ is about two times larger than $S'_j$,
both entropies have a similar $j$ dependence
with the maximum at $a=4$ ($j=2$).

\subsection{Effects of Variation in ISI }

We will discuss in this subsection, how the variation in ISI 
of the spike signal modifies the result of the WT.
Firstly we equally change both $T_{{\rm o}1}$ and $T_{{\rm o}2}$
as $T_{{\rm o}1} = T_{{\rm o}2} \equiv T_{\rm o}$.
Figure 10(a), (b), (c) and (d) 
show the analyzed spike signals (upper frames) and the energy-density 
patterns (lower frames) for $T_{\rm o}$ = 15, 25, 35 and 50 ms,
respectively.  We note that the size of patterns becomes
large almost in proportion to an increase in $T_{\rm o}$.
Figure 11(a) and (b) show the $a$ dependence of the energy-density
distribution $E(a)$ and the entropy $S(a)$ in the CWT, respectively.
The peak position moves to larger $a$ for the spike
signal with larger $T_{\rm o}$, as expected.
The relevant $j$ dependence of $E_j$ and $S_j$ in the DWT is
shown in Fig. 11(c) and (d). When $T_{\rm o}$ is smaller, $E_j$ and
$S_j$ have large magnitude for smaller $j$, which is
consistent with the result obtained in the CWT.

The FT for a signal $f(t)$ is defined by
\begin{equation}
\hat{F}(\omega)= \int {\rm d}t \; e^{-i \omega t}\; f(t).
\end{equation}
When $f(t)$ is given by a delta-function-type function:
\begin{equation}
f(t)=\sum_{n=1}^{3} \; \delta(t-t_{{\rm o}n}),
\end{equation}
which is a simplification of our $M=3$ HH spike with
firing times of $t_{{\rm o}n}$,
its FT spectrum is given by
\begin{equation}
\mid \hat{F}(\omega) \mid^2
= 2\; \sum_{n=1}^{3} \; {\rm cos}(2 \pi f/f_{{\rm o}n}) + 3.
\end{equation}
In Eq.(55)
$f=\omega/2 \pi$ is the frequency and the fundamental 
frequencies $f_{{\rm o}n}$ $(n=1-3)$ are given by
$f_{{\rm o}1}=1/T_{{\rm o}1}$, $f_{{\rm o}2}=1/T_{{\rm o}2}$ 
and $f_{{\rm o}3}=1/(T_{{\rm o}1}+T_{{\rm o}2})$. 

Figures 12(a), (b) (c) and (d) show the FT frequency spectra of the HH
spike signals for  $T_{\rm o}$ = 15, 25, 35 and 50 ms, respectively.
When $T_{\rm o}$ = 25, for example, the fundamental frequencies are
$f_{{\rm o}1}=f_{{\rm o}2}=$ 40 and $f_{{\rm o}3}$=20 Hz, 
the former being coincidentally the 
second harmonics of the latter. The dashed curve in Fig. 12(b) shows
the FT spectra given by Eq.(55) for a delta-function-type signal
given by Eq.(54) with $T_{\rm o}=25$ ms, which is in good
agreement with the FT spectra of the HH spike at $f < 150$ Hz.

Next we investigate the effect of ISI on the WT result
by changing $T_{{\rm o}1}$ but keeping 
$T_{{\rm o}1}+T_{{\rm o}2}$= 50 ms.
The analyzed spike signals (upper frames) 
and the pattern of the energy-density
distribution (lower frames) for $T_{{\rm o}1}$ = 10, 15, 20 and 25 ms
are shown in Fig. 13(a), (b), (c) and (d), respectively.
Note that the average ISIs are the same (25 ms)
but their RMS are different:
$(<(T_{\rm on}^2-<T_{\rm on}>^2)>)^{1/2}$ = 15, 10, 5 and 0 ms
for $T_{{\rm o}1}$ = 10, 15, 20 and 25 ms, respectively.
For the case of $T_{{\rm o}1} (= T_{{\rm o}2}$) = 25 ms, 
the calculated pattern
in Fig. 13(d) is nearly symmetric with respect
to the axis of
$b$ = 128 ms, at which the second spike fires.
With introducing a small asymmetry in ISI: 
$T_{{\rm o}1}$ = 20 ($\;T_{{\rm o}2}$ = 30) ms,
the pattern becomes asymmetric, as shown in Fig. 13(c).
This is more clearly seen in Fig. 13(a) for the case of 
$T_{{\rm o}1}$ = 10 ($\;T_{{\rm o}2}$ = 40) ms: 
smaller $T_{{\rm o}1}$ and larger $\;T_{{\rm o}2}$
yield smaller and  larger patterns, respectively.
Figure  14(a) and (b) show the $a$ dependence of
$E(a)$ and $S(a)$, respectively.
In the case of $T_{o1}$ = 25 ms, $E(a)$ has a single maximum
at $a \sim 20$. On the contrary, in the case of $T_{{\rm o}1}$ = 10 ms,
$E(a)$ and $S(a)$ have double maxima: the peak
at smaller $a$ arises from the contribution due to
a smaller $T_{{\rm o}1}$ and
the peak at larger $a$ from a larger $T_{{\rm o}2}$.
These features are consistent with the results obtained
in the DWT, which are shown in Figs. 14(c) and 14(d).
Figures 13 and 14 clearly show that the response of the HH neuron
cannot described only by the rate of the 
input pulses \cite{Hasegawa00},
in agreement with experiments \cite{Segundo63}\cite{Mainen95}.

Figure 15 shows the FT spectra for spike signals
for $T_{{\rm o}1}$ = 10, 15, 20 and 25 ms.
In the case of $T_{{\rm o}1}$ = 15 ms, for example,
the fundamental frequencies
are 66.7, 28.6 and 20 Hz.
Although a glimpse of Fig. 15(b) suggests that fundamental
frequencies are 60 and 80 Hz, it is not true.
The dashed curve in Fig. 15(b) shows the FT spectra
given by Eq.(55) for a delta-function signal given by
Eq.(54) with $f_{{\rm o}1}=66.7$, $f_{{\rm o}2}=28.6$
and $f_{{\rm o}3}=20$ Hz, which is in good agreement
with the FT spectra of HH spikes at $f < 150$ Hz.

\subsection{Effects of Noises}

We will study, in this subsection, the effect of OU noises
given by Eqs.(7)-(9).
Figure 16(a), (b), (c) and (d) show the spike signals 
(upper frames)
and the CWT patterns of the energy-density distribution
(lower frames) for $D$ = 0, 1, 2 and 3, respectively.
The result without noises ($D = 0$) has been shown 
previously [e.g. Fig. 4(b)].
When noises are introduced, spike signals 
lead to fine structures
in the CWT patterns.
For a large $D = 3$, applied noises trigger a spurious 
spike at $t = 203$ ms.

Figure 17(a) and (b) show $E(a)$ and $S_1(a)$ for
$D =$ 0, 1, 2 and 3 calculated by the CWT coefficients.
Applied noises provide extra energy contributions in
a fairly wide range of $a$ value. Particularly for $D = 3$,
$E(a)$ and $S(a)$ are much increased by
a spuriously triggered spike at $t = 203$ ms.

Similar results are obtained in the $j$-dependent energy ($E_j$)
and entropy ($S_j$) calculated by the DWT, which are plotted
in Fig. 17(c) and (d).

Spike signals with noises are analyzed also by the FT, whose
frequency spectra for $D =$ 0, 1, 2 and 3 are plotted
in Fig. 18(a), (b), (c) and (d), respectively.
It is interesting that peaks of the fundamental frequency
of 40 Hz and its harmonics are sharpened by noises.

\vspace{0.5cm}
\section{Conclusion and Discussion}

One of the advantages of the 
DWT is that we can easily separate noise components
from a given signal.
As the first example, we discuss the denoising 
by using the DWT.\cite{Bartnik92}-\cite{Quiroga00}
The key point in the denoising is how to choose which wavelet 
coefficients are correlated with the signal and which ones with noises.
The simple
denoising is to neglect some DWT expansion coefficients 
when reproducing the signal by
the inverse wavelet transformation.
In getting the inverse WT by
\begin{equation}
f_{\rm I}(t) = \sum_j \;\sum_k c_{jk}^{\rm dn} \;\psi_{jk}(t),
\end{equation}
we may assume that the components for $a < a_{\rm c}$ in the $(a, b)$ plane
arise from noises to set the denoising coefficients $c_{jk}^{\rm dn}$ as
\begin{eqnarray}
c_{jk}^{\rm dn} & = & c_{jk}, \;\;\;
\mbox{for $j \geq j_{\rm c}$ ($a \geq a_{\rm c}$)} \nonumber \\
& = & 0,  \;\;\;\;\; \mbox{otherwise}
\;\;\;\;\;\;\;\;\;\;\;\;\;\;\;\;\;\;\;\;\;\;  
\;\;\;\;\;\;\;\;\;\;\;\;\;\;\;\;\;\;\;\;\;\;  
\mbox{(Method I)}
\end{eqnarray}
where $j_c = {\rm log}_2\; a_{\rm c}$ is the critical $j$ value.
A demonstration of denoising of spike signals 
with $M=3$ and $D=3$ is given in Fig. 19.
Figure 19(a) expresses the original output spike, and Fig. 19(b)
shows the inverse signals by using the method I 
[eq.(4.2)] with $j_{\rm c}=2$.
For a comparison, we show the relevant results 
for the postsynaptic input $I_{\rm i}$:
the original signal in Fig. 19(d) and
the inverse signal with denoising for $j_{\rm c}=2$
in Fig. 19(e).

Since the DWT transforms the original signal to 
the two-dimensional ($a,b$) plane, it has much freedom
than the FT, which makes it possible for us
to adopt the more sophisticated denoising.
We may assume that the components for 
$b < b_{\rm L}$ or $b > b_{\rm U}$ 
at $a < a_{\rm c}$ in the (a,b) plane
are noises to set the denoising WT coefficients as
\begin{eqnarray}
c_{jk}^{\rm dn} & = & c_{jk}, \;\;\;
\mbox{for $j \leq j_{\rm c}$ 
or $k_{\rm L} \leq k \leq k_{\rm U}$} \nonumber \\
& = & 0,  \;\;\;\;\; \mbox{otherwise}
\;\;\;\;\;\;\;\;\;\;\;\;\;\;\;\;\;\;\;\;\;\;  
\;\;\;\;\;\;\;\;\;\;\;\;\;\;\;\;\;\;\;\;\;\;  
\mbox{(Method II)}
\end{eqnarray}
where $j_{\rm c} = {\rm log}_2\; a_{\rm c}$, 
and $k_{\rm L}=b_{\rm L} \; 2^{-j}$ 
and $k_{\rm U}= b_{\rm U} \; 2^{-j}$ are 
lower and upper critical $k$ values.
Figures 19(c) and 19 (f) show the donoising signals
of the output and postsynaptic signals, respectively,
by using the method II [Eq.(4.3)]
with $b_{\rm L}=40$, $b_{\rm U}=220$ and $j_{\rm c}=4$.
The denoising method II given by Eq. (4.3)
is better than the method I given by eq.(4.2)
because the magnitude of a spurious spike at $t \sim 200$ ms
is much reduced while the signal at $100 < t < 160$ ms is
better reproduced in the method II than in the method I.

As the second example using the DWT, we discuss
the signal-to-noise ratio (SNR).
From the above consideration, we may define
the signal component $A_{\rm s}$ and 
the noise component $A_{\rm n}$ by
\begin{equation}
A_{\rm s} = \sum_j \sum_k \mid c_{jk}^{\rm dn} \mid^2,
\end{equation}
\begin{equation}
A_{\rm n} = \sum_j \sum_k (\mid c_{jk} 
\mid^2 - \mid c_{jk}^{\rm dn} \mid^2)
\end{equation}
which lead the SNR $\eta$:
\begin{equation}
\eta = {\rm log}_{10} (A_{\rm s}/A_{\rm n}).
\end{equation}
The solid curves in Fig. 20(a) and Fig. 20(b) shows
the $D$-dependent SNR of output (V) and 
input signals ($I_i$),
respectively,
with $M=3$ calculated by the denoising method I
given by eq.(4.2) with $j_{\rm c}=2$.
It might be strange that SNR is finite even for no noises ($D=0$).
This is because signals for $D=0$ already included in
the $j=1$ component are regarded as noises in the denoising method I 
though its magnitude is small. On the contrary,
the dashed curves in Fig. 20(a) and 20(b) show the SNR
as a function of $D$ calculated by the denoisng
method II given by eq.(4.3)
with $b_{\rm L}=40$, $b_{\rm U}=220$ and $j_{\rm c}=4$.
Because of the condition imposed for 
the $a$ variable in the method II,
the noise contribution is much reduced for small $D$, yielding
a large SNR.
With increasing the value of $D$, SNR of the input
and output signals decreases as expected.

Finally, 
we apply the CWT to a signal which is a simplified one of 
the original HH neuron spike,
in order to get some insight how the detailed shape
of spikes is relevant to information transmission. 
As its candidate, we adopt a square pulse with 
the width of 1 ms, the delta-function-type signal.
Figures 21(a) and (b) show the
analyzed signal and its energy-density 
distribution in the CWT, respectively.
Comparing Fig. 21(b) with Fig. 4 (b) for the original HH spike,
we note that except for very small value of $a (\siml 4)$, the 
CWT pattern for the delta-function-type signal is almost the same
as that of the original HH spike.
This suggests that information is predominantly 
carried by firing times of 
spikes, but not by their detailed structure. This 
agrees with the conventional wisdom
in the neuroscience community.


To summarize, we have analyzed transient spike signals with the use 
of both the CWT and DWT. 
The WT has been shown to be more useful than the FT for an analysis
of transient signals like spikes.
The WT now has a wide-range of applications
including the information compression like MPEG and JPEG.
It might be possible that real neural systems
adopt the WT-like technique for their efficient information
transaction.

\section*{Acknowledgements}
This work is partly supported by
a Grant-in-Aid for Scientific Research from the Japanese 
Ministry of Education, Culture, Sports, Science and Technology.

\newpage
%

%

\newpage

\vspace{0.5cm}

\noindent{\large\bf  Figure Captions}   \par

\vspace{0.5cm}

\noindent
{\bf Fig. 1} 
The time course of (a) the input potential $U_{\rm i}$,
(b) the postsynaptic potential $I$ ($=I_{\rm ps}+I_{\rm n}$)
and (c) the membrane potential $V$.   
\vspace{0.5cm}

\noindent
{\bf Fig. 2}
(a) The analyzed signal, (b) the pattern of the CWT coefficients
$c(a, b)$, and (c) that of the energy-density distribution 
$\epsilon(a, b) \equiv \mid c(a, b) \mid^2$.
Scales of colors from minimum to maximum in (b) and (c) are shown.
\vspace{0.5cm}

\noindent
{\bf Fig. 3}
The $b$ dependence of the CWT coefficients $c(a, b)$
for specified $a$ values.
\vspace{0.5cm}

\noindent
{\bf Fig. 4}
(a) The $M=3$ spike signal, (b) the profile of energy distribution
$\epsilon(a, b)$ analyzed by the CWT, and (c) that analyzed by the DWT.
\vspace{0.5cm}

\noindent
{\bf Fig. 5}
The $b$ dependence of the CWT coefficients $c(a,b)$
for specified $a$ values.
\vspace{0.5cm}

\noindent
{\bf Fig. 6} 
(a) The $a$-dependent energy [$E(a)$ and $E(a)/a^2$]
and (b) entropy [$S(a)$ and $S(a)/a^2$] in CWT.
(c) The $j$-dependent energy ($E_j$) and 
(d) entropy ($S_j$, $S'_j$) in DWT,
solid lines being drawn only for a guide of eye (see text).
\vspace{0.5cm}

\noindent
{\bf Fig. 7} 
(a) The $a$-dependent average of $c(a ,b)$
[$\mu(a)$ and $\mu(a)/a^2$], and 
(b) their RMS [$\sigma(a)$ and $\sigma/a^2$] in CWT.
(c) The $j$-dependent average of $c_{jk}$
($\mu_j$), and
(d) their RMS ($\sigma_j$) in DWT,
solid lines being drawn only for a guide of eye (see text).
\vspace{0.5cm}

\noindent
{\bf Fig. 8}
The DWT decomposition of the signal to various components;
$V = \sum_{j=1}^6 d_j + a_6$.  
\vspace{0.5cm}

\noindent
{\bf Fig. 9} 
(a) The contour map of the entropy distribution $s(a, b)$.
(b) The $a$ dependence of $s(a, b)$ 
for $b$ = 103 (solid curve), 109 (dot-dashed curve) 
and 115 ms (dashed curve), 
whose positions are indicated by vertical, dashed lines in (a).
(c) The $b$ dependence of $s(a, b)$ 
for $a$ = 10 (solid curve), 20 (dot-dashed curve) 
and 30 (dashed curve), 
whose positions are indicated by horizontal, dashed lines in (a).

\vspace{0.5cm}

\noindent
{\bf Fig. 10}
The time dependence of spikes (upper frames)
and the energy-density patterns (lower frames)
in the CWT for (a) $T_{\rm o}$ = 15, (b) 25, (c) 35 and (d) 50 ms.   

\vspace{0.5cm}

\noindent
{\bf Fig. 11}
(a) The $a$ dependence of $E(a)$ and (b) the entropy $S_1(a)$
in the CWT for $T_{\rm o}$ = 15, 25, 35 and 50 ms.
(c) The $j$ dependence of $E_j$ and (d) the entropy $S_j$
in the DWT for $T_{\rm o}$ = 15, 25, 35 and 50 ms.

\vspace{0.5cm}
\noindent
{\bf Fig. 12} 
The FT spectra for  
(a) $T_{\rm o}$ = 15, (b) 25, (c) 35 and (d) 50 ms.   
The dashed curve in (b) expresses the FT spectra
given by Eq.(55) for a delta-function-type
signal given by Eq.(54) with $T_{\rm o}=25$ ms.
\vspace{0.5cm}

\noindent
{\bf Fig. 13}
The time dependence of spikes (upper frames)
and the energy-density patterns (lower frames)
in the CWT for (a) $T_{{\rm o}1}$ = 10, (b) 15, (c) 20 and (d) 25 ms.   

\vspace{0.5cm}

\noindent
{\bf Fig. 14}
(a) The $a$ dependence of $E(a)$ and (b) the entropy $S_1(a)$
in the CWT for $T_{{\rm o}1}$ = 10, 15, 20 and 25 ms.
(c) The $j$ dependence of $E_j$ and (d) the entropy $S_J$
in the DWT for $T_{{\rm o}1}$ = 10, 15, 20 and 25 ms.

\vspace{0.5cm}

\noindent
{\bf Fig. 15} 
The FT spectra for  
(a) $T_{{\rm o}1}$ = 10, (b) 15, (c) 20 and (d) 25 ms.
The dashed curve in (b) expresses the FT spectra
given by Eq.(55) for a delta-function-type
signal given by Eq.(54) with $T_{{\rm o}1}=15$ ms.

\vspace{0.5cm}

\noindent
{\bf Fig. 16}
The time dependence of spikes (upper frames)
and the energy-density patterns (lower frames)
in the CWT
for (a) $D$ = 0, (b) 1, (c) 2, and (d) 3.   

\vspace{0.5cm}

\noindent
{\bf Fig. 17}
(a) The $a$ dependence of $E(a)$ and (b) the entropy $S_1(a)$
in the CWT for $D$ = 0, 1, 2 and 3.
(c) The $j$ dependence of $E_j$ and (d) the entropy $S_j$
in the DWT for $D$ = 0, 1, 2 and 3.

\vspace{0.5cm}

\noindent
{\bf Fig. 18} 
The FT spectra for  
(a) $D$ = 0, (b) 1, (c) 2, and (d) 3.

\vspace{0.5cm}

\noindent
{\bf Fig. 19}
(a) The original output spike $V$ with $D=3$,
(b) its inverse signal
by the denoising method I [eq.(4.2)] with $j_{\rm c}=2$, 
and (c) that by the denoising method II [eq.(4.3)]
with $b_{\rm L}=40$, $b_{\rm U}=220$ and $j_{\rm c}=4$.
(d) The original input signal $I_i$ with $D=3$
(e) its inverse signal
by the denoising method I [eq.(4.2)] with $j_{\rm c}=2$, 
and (f) that by the denoising method II [eq.(4.3)]
with $b_{\rm L}=40$, $b_{\rm U}=220$ and $j_{\rm c}=4$.

\vspace{0.5cm}

\noindent
{\bf Fig. 20}
The $D$ dependence of SNR
of (a) the output spike
signal $V$ and (b) the input presynaptic signal $I_{\rm i}$;
the solid curves show the results by using 
the denoising method I [eq.(4.2)] with $j_{\rm c}=2$, 
and the dashed curves express the
results by using the 
denoising method II [eq.(4.3)]
with $b_{\rm L}=40$, $b_{\rm U}=220$ 
and $j_{\rm c}=4$ (see text).

\vspace{0.5cm}

\noindent
{\bf Fig. 21}
(a) The analyzed delta-function-type
signal and (b) the energy-density pattern
in the CWT (see text).

\end{document}